# Quantitatively Connecting Experimental Time-Temperature-Superposition-Breakdown of Polymers near the Glass Transition to Dynamic Heterogeneity via the Heterogeneous Rouse Model

Peijing Yue, David S. Simmons*

Department of Chemical, Biological, and Materials Engineering, University of South Florida, Tampa, Florida 33612, USA.

**ABSTRACT:** Polymers near the glass transition temperature $T_g$ often exhibit a breakdown of time-temperature-superposition (TTS), with chain relaxation times and viscosity exhibiting a weaker temperature dependence than segmental relaxation times. The origin of this onset of thermorheological complexity has remained unsettled and a matter of debate. Here we extend the Heterogeneous Rouse Model (HRM), which generalizes the Rouse model to account for dynamic heterogeneity, to make predictions for the relaxation modulus $G(t)$ and complex modulus $G^*(\omega)$ of unentangled polymers near $T_g$. The HRM predicts that $G(t)$ and $G^*(\omega)$ exhibit enhanced effective scaling exponents in the Rouse regime in the presence of dynamic heterogeneity, with a more rapid decay from the glassy plateau emerging as the system becomes more dynamically heterogeneous on cooling. This behavior is predicted to emerge from a strand-length dependence of the *moment* of the segmental mobility distribution probed by chain dynamics. We show that the HRM predictions are in good accord with experimental complex modulus data for polystyrene, poly(methyl methacrylate), and poly(2-vinyl pyridine). The HRM also predicts the onset of distinct temperature dependences among chain scale quantities such as terminal relaxation time and viscosity in our experimental systems, apparently resolving one of the most significant standing objections to a heterogeneity-based origin of TTS-breakdown. The HRM thus provides a generalized theory of the chain-scale linear rheological response of unentangled polymers near $T_g$, accounting for the origin of TTS-breakdown at a molecular mechanistic level. It also points towards a new strategy of inferring the dynamic heterogeneity of glass-forming polymeric systems based on the temperature-evolution of modified scaling in the Rouse regime.

## Introduction

The time temperature superposition (TTS) principle is the practical underpinning of analysis of temperature- and frequency-dependent linear-regime dynamics in diverse polymers. This principle is grounded in the proposition that distinct polymer relaxation modes, ranging from segmental to whole-chain, are governed by the same underlying molecular friction mechanism and therefore share a common temperature dependence. This proposition is most fundamentally rooted in the Rouse model, which provides a theoretical argument for a temperature-invariant proportionality between segmental and chain dynamics[1]. However, pioneering work by Plazek and coworkers in the 1960's,[2] and since followed by numerous additional studies[3–15], demonstrates that TTS breaks down in many polymers near $T_g$, signaling an onset of thermorheological complexity. The fundamental origin of this breakdown remains an open question and a matter of considerable debate[10,16,17]. This question is both fundamentally relevant to the understanding of polymer dynamics and practically relevant to characterization and quantification of dynamics in diverse polymers.

TTS breakdown in near-$T_g$ polymers is characterized by a decoupling of the temperature dependences of segmental and chain normal mode relaxations. Typically, segmental relaxation time $\tau_o$ and chain relaxation time $\tau_n$ are observed to obey a fractional power law decoupling relation[10],

$$\frac{\tau_0}{\tau_n} = a \tau_n^{-\varepsilon} \qquad (1)$$

where $a$ is a constant of proportionality and $\varepsilon$ is called the decoupling exponent, which quantifies the extent to which chain and segmental dynamics obey distinct temperature dependences. The magnitude of this effect varies considerably between polymers, for example with $\varepsilon > 0.5$ for polycarbonate and in the vicinity of 0.2 for polyisoprene[10]. Experimental evidence suggests that the decoupling emerges in the Rouse regime: (1) similar effects are observed in both entangled and unentangled polymers[10]; (2) relaxation spectra in the TTS regime exhibit deviations from classical Rouse scaling in the nominally Rouse regime[2,18].

There has been considerable debate over the origins of this effect. A series of papers have proposed that it can be explained on the basis of the Coupling Model of Ngai and coworkers[6–8,19], which posits that distinct relaxation processes involve distinct degrees of 'coupling' to their molecular environment. Inoue et al. suggested that, as a phenomenological matter, data in the TTS-breakdown regime could be treated by assuming distinct temperature dependences for the glassy relaxation and 'rubbery relaxation', without proposing a precise molecular polymeric physical model for this effect[18,20]. Alternatively, Lin suggested that this TTS breakdown can be explained via the Extended Reptation Theory, based on the proposition that segmental and chain modes differentially sample entropic and energetic relaxation barriers with distinct temperature dependences[21]. In another approach, Loring proposed a model in which TTS-breakdown and deviations from Rouse scaling emerge from temporal fluctuations in the obstacles imposed on a segment by surrounding segments[22,23].

In 2009, Sokolov and Schweizer proposed that chain-segment decoupling may result from averaging over emergent spatially heterogeneous dynamics upon cooling towards the glass transition temperature $T_g$.[10] The attractiveness of this proposition lies in the fact that dynamic heterogeneity is a well-established phenomenological feature of glass-forming liquids[24–40]. This proposition shares with the work of Loring[22,23] the idea that dynamical heterogeneity plays a central role in thermorheological complexity near $T_g$, but Sokolov and Schweizer's proposed conceptual focus is on the effects of *averaging* over this heterogeneity rather than on the effect of temporal fluctuations of local obstacles to relaxation (as in Loring's case). The basic premise of this proposed mechanism is that segmental and chain modes reflect distinct moments of an underlying distribution of relaxation times. Differential averaging over heterogeneous dynamics has also long been proposed as an explanation for other near-$T_g$ decouplings of distinct relaxation processes, such as viscous relaxation and diffusion[41,42]. This latter canonical example is referred to as Stokes-Einstein breakdown.

However, Ngai, Plazek, and Roland have raised several objections to the heterogeneity picture of chain normal mode decoupling[16]. Most significantly, the heterogeneity scenario requires that the breadth of the distribution of local mobilities *grows* on cooling, as it is this growth that is proposed to drive thermal decoupling. There is appreciable evidence for this proposition in the literature[27,30,31,40,43–45]. However, Ngai et al. argue that several small-molecule systems exhibit Stokes-Einstein breakdown despite displaying a lack of temperature dependence in the breadth of dielectric relaxation peaks[16]. Based on the common belief that the breadth of dielectric relaxation peaks reports on the breadth of the underlying distribution of relaxation times, they suggest that this feature of the dielectric spectroscopy points towards a temperature-invariant relaxation spectrum in these systems, contradicting the proposition that decoupling can be generally connected to a growing dynamic heterogeneity on cooling. Second, they argue that the observation of decoupled viscosity and chain diffusion in some polymers[5] indicates that spatial averaging over dynamic heterogeneity cannot account for chain normal mode decoupling, since both of these processes take place at the chain scale.

Our group has recently suggested a potential resolution to the first objection above – the proposition that the stretching exponent (or breadth of the relaxation beak) probed by dielectric spectroscopy may not quantitatively report on the breadth of the underlying distribution of relaxation times[46]. Several pieces of evidence support this viewpoint. First, simulations have demonstrated that stretching exponents associated with reorientational relaxation (the main relaxation mode probed by dielectric spectroscopy in the frequency range relevant to decoupling) and those associated with translational relaxation can exhibit qualitatively different values and temperature dependences[27,46]. This finding speaks against the proposition that deviations from exponential relaxation can report on a single underlying relaxation time distribution in a general way. Second, recent simulations in the isoconfigurational ensemble have demonstrated that deviations from exponential relaxation do not emerge purely from spatial averaging, but also involve a significant local component of nonexponentiality, with the relative roles of spatial averaging and local nonexponentiality depending on chemistry and relaxation function probed[25,27]. These findings again indicate that the breadth of relaxation peaks in dielectric spectroscopy measurements cannot generally be interpreted as a quantitative measure of the microscopic relaxation time distribution. More experimental validation of this proposition is still needed before it can be considered settled. However, these early results at least point to a potential resolution of the presence of decoupling in systems lacking a temperature dependence of the breadth of dielectric relaxation spectra: this breadth may simply not be perfectly correlated with the underlying extent of dynamical heterogeneity.

In response to the second objection above, Sokolov and Schweizer argued that the observed apparent viscosity-chain diffusion decoupling was in entangled polymers far above $T_g$, and may involve entanglement physics totally separate from the near-$T_g$ TTS-breakdown problem[17]. However, even if one disregards this potential difference, it is presently unclear whether a heterogeneity-based scenario for TTS-breakdown must genuinely lead to coupled chain viscosity and diffusion. A resolution to this question requires a more quantitative description of polymer chain dynamics in the presence of dynamical heterogeneity. More broadly, a full assessment of the merits of the proposed heterogeneity mechanism for TTS breakdown requires a theory capable of predicting linear polymer viscoelastic response functions in the presence of dynamical heterogeneity.

Inspired by Sokolov and Schweizer's proposal, and seeking to provide the required more quantitative description of polymer dynamics within the heterogeneous setting known to characterize polymers near $T_g$, in 2018 we developed the "Heterogeneous Rouse Model" (HRM).[46] The HRM generalizes the Rouse model to account for the presence of a distribution of segmental relaxation times in polymers near $T_g$.[46] There is precedent for considering the role of heterogeneity in Rouse dynamics: Rubinstein and Colby developed a Rouse model of constraint release in long polymer chains (with 'beads' corresponded to entanglements rather than repeat units), wherein a distribution of 'bead' mobilities reflects variations in the proximity of entanglements to chain ends[47]. As discussed above, Loring also considered the role of temporal fluctuations in local obstacles to segmental motion within a Rouse-like picture[22,23]. In the HRM, we instead introduce heterogeneity at the level of *spatial* variations in *individual chain segment mobility*, and we employ this model to predict deviations from Rouse scaling in the TTS-breakdown regime.

In our prior work, we demonstrated that this theory predicts breakdown in the Rouse scaling of strand friction coefficients at low temperature in a simulated bead-spring polymer chain, with only a single temperature-independent adjustable timescale prefactor implicated in the prediction[46]. These findings demonstrated that temperature-dependent heterogeneity can yield chain normal mode decoupling in a model polymer. However, they left open the question of the implications of this model for rheological response functions in experimental systems.

Here, we derive the implications of the HRM for rheological response spectra in the TTS-breakdown regime. We then report on new measurements of relaxation spectra in several experimental polymers exhibiting TTS-breakdown near $T_g$, and we show that these spectra are well-described and the breakdown well-explained by the HRM. These findings provide new empirical support for the HRM scenario wherein near-$T_g$ TTS-breakdown and chain normal mode decoupling emanate from emergent dynamic heterogeneity. Finally, we show that this theory can explain observed decouplings between distinct chain-scale dynamical quantities. This resolves the second quandary raised by Ngai and coworkers described above, by demonstrating that a heterogeneous origin of chain normal mode decoupling is consistent with observed decoupling between nominally chain-scale quantities.

The HRM is at present limited to non-entangled systems, and further extension to incorporate topological entanglements (effectively to develop a heterogeneous reptation theory) is needed for a full description of long chain dynamics in the TTS-breakdown regime. However, TTS-breakdown and chain normal mode decoupling are reported in both subentangled and entangled systems with little or no qualitative difference between the two cases[10], suggesting that this distinction is not qualitatively important to the underlying mechanism of chain normal mode decoupling. Indeed, the HRM foreshadows this finding, predicting that much of the heterogeneity should commonly be 'averaged out' by the time the chain samples the entanglement strand via Rouse motion, such that longer-time reptation may in many cases sample an effectively near-homogenous environment on the scale of the entanglement strand.

## Theory

### Derivation

We begin by reviewing the essential elements of the HRM as previously developed at the level of chain friction coefficients only.[46] The theory constitutes a generalization of the Rouse model, which in its standard form assumes that the segmental friction coefficient is monodisperse. Specifically, classical Rouse theory computes the chain friction coefficient as a sum over segmental friction coefficients, leading to the proposition that the segmental friction coefficient $\zeta_o$ and chain friction coefficient $\zeta_n$ are related as $\zeta_n = n\zeta_o$. The HRM relaxes this assumption, instead beginning with a description of chain diffusion within the presence of a distribution of segmental mobilities. This leads to the prediction that the chain friction coefficient is related to the segmental friction coefficient as[46]

$$\langle \zeta_n \rangle = n \langle \zeta_0 \rangle \left\langle \frac{1/\langle \zeta_0 \rangle}{\langle 1/\zeta_0 \rangle_n} \right\rangle_{N/n} \tag{2}$$

Here $\langle\zeta_o\rangle$ is the segmental friction coefficient linearly averaged over all $N$ segments in the system. $\langle 1/\zeta_o\rangle_n$ is the inverse segmental friction coefficient averaged over the $n$ segments in an $n$-mer; since $n$ is generally finite, this is not a single number but a distribution if $\zeta_o$ is itself a distributed quantity. The outer brackets with a subscript of $N/n$ denote an average of this quantity over all $N/n$ strands of length $n$ in the system. The leading factor of $n\langle\zeta_o\rangle$ is the Rouse prediction; the final factor is thus a correction factor to Rouse. This equation indicates that the friction coefficient interpolates between reflecting the first moment of the friction coefficient distribution (at the near-segment level) and reflecting one over the inverse moment of the distribution (for long strands). Physically, this can be understood as reflecting Stokes-Einstein breakdown at a monomeric level, with progressive 'pre-averaging' of segmental dynamical heterogeneity mitigating this breakdown in progressively lower chain modes.

Equation (2) is grounded in the underlying physics of the Rouse theory, while simply relaxing the Rouse assumption of uniform friction coefficients. The additional approximations introduced are relatively well controlled: the central limit theorem is employed to quantify the manner in which chain segments sample over heterogeneity; segmental motion is modeled as obeying a Gaussian path process. The central limit theorem is rigorous only in the limit of large $n$; however, since the underlying path process of segments is itself Gaussian to leading order[48,49], corrections to this leading order description are expected (and confirmed by comparison to simulation[46]) to be small, and admit the possibility of systematic introduction of higher order terms. Similarly, deviations from Gaussian motion can be quantified and in principle corrected for[49], but results from our prior work indicate that these corrections are small enough so as not to appreciably perturb results[46]. The HRM thus involves fewer assumptions than the Rouse model, and introduces only a pair of small and well-controlled approximations. Moreover, equation (2) involves no assumptions regarding the *form* of the distribution of segmental mobilities or friction coefficients.

In order to proceed further, the HRM theory employs a log-normal distribution of segmental mobilities. This choice is motivated both by theoretical considerations[50,51] and by the empirical observation in simulations[27,52] and experiments[26] that local relaxation times and other dynamical properties in glass-forming liquids are to leading order log-normally distributed. This distribution also corresponds with a Gaussian distribution of the activation barrier to segmental relaxation.

The central working prediction emerging from the combination of equation (2) with a log-normal distribution is that the ratio of the strand to segmental friction coefficients is given by[46]

$$\frac{\langle \zeta_n \rangle}{\langle \zeta_0 \rangle} = n\left[1 - \left(1 - \frac{1}{n}\right)\left(1 - \frac{1}{r_{SE}}\right)\right] \quad (3)$$

Here $r_{SE}$ is the Stokes-Einstein breakdown ratio, given by

$$r_{SE} \equiv \frac{\langle \zeta_0 \rangle}{\langle \zeta_0 \rangle_{SE}} = \frac{\langle 1/D_0 \rangle}{1/\langle D_0 \rangle}, \quad (4)$$

which is the ratio of the actual mean system *monomeric* friction coefficient to the value of the monomeric friction coefficient predicted from the monomeric mobility via the Einstein relation. Within the HRM, this ratio is directly related to the log-normal breadth parameter controlling the distribution-width of monomeric diffusivities,

$$r_{SE} = \exp\left[\sigma^2\right], \quad (5)$$

where $\sigma$ is the breadth parameter of the log-normal distribution of local mobilities. $r_{se}$ is thus a measure of the breadth of the segmental mobility distribution in the polymer (i.e. the amount of dynamic heterogeneity), which is expected to grow on cooling. In our prior work, we demonstrated that equation (3) *quantitatively* describes deviations from Rouse scaling of translational friction coefficients in molecular dynamics simulations of model bead-spring polymers near $T_g$.[46]

We now extend this model to predict linear rheological response functions and to compare them to experimental measurements of the same. To do so, we first extend from friction coefficients to molecular relaxation times by introducing to equation (3) the extra product of $n$ associated with the motion of a Gaussian chain over its own square gyration radius $\langle R_g^2 \rangle$:

$$\frac{\langle \tau_n \rangle}{\langle \tau_0 \rangle} = n^2\left[1 - \left(1 - \frac{1}{n}\right)\left(1 - \frac{1}{r_{SE}}\right)\right] \quad (6)$$

To compute a relaxation modulus from this relaxation time, we employ the continuous spectrum scaling approach outlined by Rubinstein and Colby[1]: we solve equation (6) for the smallest unrelaxed strand length $n$ at a given time $t$ and then employ the observation that the modulus of the fluid is equal to $k_BT$ times the density of unrelaxed strands. Enforcing agreement with the pre-

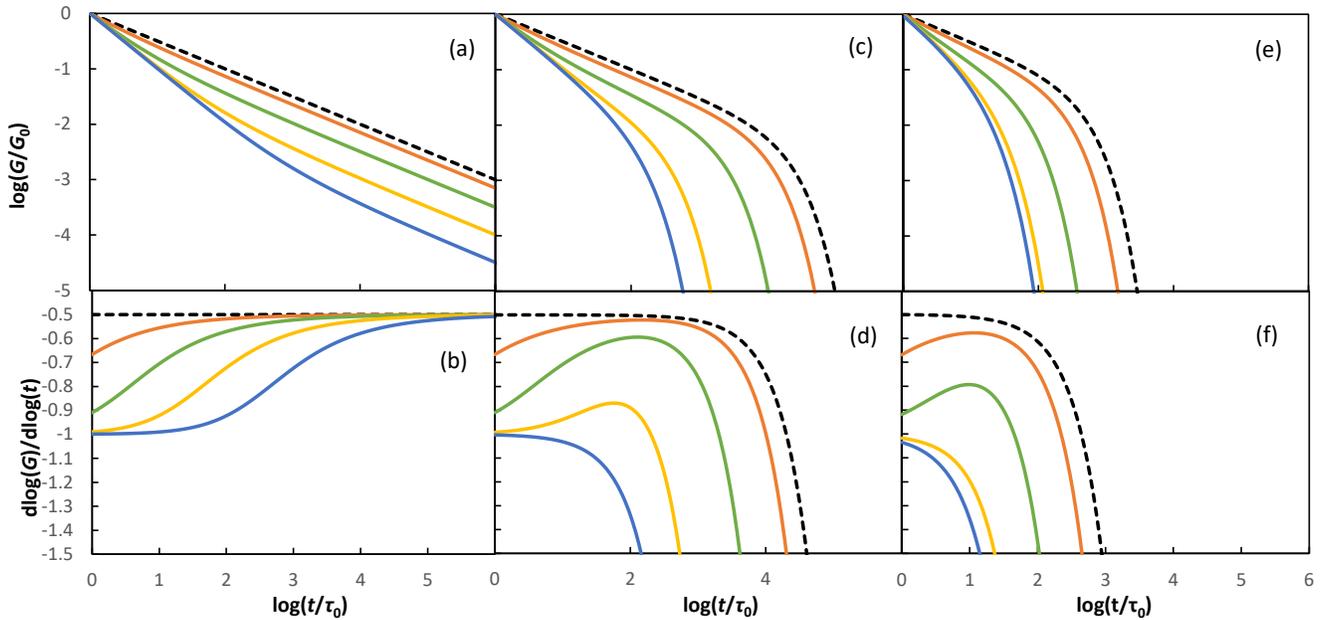

Figure 1. (a, c, and e) log-log plot of HRM predictions of $G(t)/G_0$ vs $t/t_0$, for hypothetical unentangled chains with (a) $n \to \infty$, (c) $n = 200$, and (e) $n = 30$. From top to bottom, curves correspond to $r_{se} = 1$ (Rouse model, dashed black line), 2 (orange), 10 (green), 100 (yellow), and 1000 (blue). (b, d and f) corresponding logarithmic slopes for the same data sets.

Rouse modulus $G_0$ at $t = \langle\tau_0\rangle$ gives for the relaxation modulus

$$\frac{G(t)}{G_0} = 2\left[1 - r_{SE} + \sqrt{1 + 2r_{SE}\left(2\frac{t}{\tau_0} - 1\right) + r_{SE}^2}\right]^{-1} \quad \tau_0 < t < \tau_N \quad (7)$$

Further introducing the terminal relaxation of the chain, and approximating the modulus for times less than $\tau_0$ as $G_0$ (we note that this approximation neglects details of the segmental relaxation process itself), yields an HRM prediction for the relaxation modulus at all times:

$$\frac{G(t)}{G_0} = \begin{cases} 1 & \tilde{t} < 1 \\ 2\left[\dfrac{1 - r_{SE} +}{\sqrt{(1-r_{SE})^2 + 4r_{SE}\tilde{t}}}\right]^{-1} \exp\left(-\dfrac{\tilde{t}}{n}\dfrac{r_{SE}}{r_{SE} - 1 + n}\right) & \tilde{t} > 1 \end{cases} \quad (8)$$

where the dimensionless time $\tilde{t} = t/\tau_0$. Notably, when $r_{SE}$ goes to one (i.e. in the absence of dynamic heterogeneity), equation (8) reduces to the classical Rouse model.

$$\frac{G(t)}{G_0} = \begin{cases} 1 & \tilde{t} < 1 \\ \dfrac{1}{\sqrt{\tilde{t}}}\exp\left(-\dfrac{\tilde{t}}{N^2}\right) & \tilde{t} > 1 \end{cases} \quad (9)$$

We then can compute a complex modulus from equation (8) via a Fourier transform. Unlike in the case of the classical Rouse model, this transform is not analytically tractable and we thus perform it numerically. The viscosity is computed via a numerical time integral over $G(t)$ in the usual way[53].

**Theoretical Predictions**

In Figure 1a, we plot the normalized relaxation modulus prediction given by equation (8) vs time, on a log-log plot, for a range of values of $r_{SE}$, in the hypothetical limit that $n \to \infty$ without chain entanglement. In Figure 1b, we plot the differential $d\log(G/G_0)/d\log(t)$, which corresponds to a plot of the slope in Figure 1a and to the effective scaling exponent $\gamma(t)$ in a relation $G(t) \sim t^{-\gamma(t)}$. As illustrated by these figures, the HRM relaxation modulus indeed recovers the Rouse prediction of $G(t) \sim t^{-1/2}$ in the limit of $r_{SE} \to 1$ (a monodisperse monomeric friction coefficient). With progressive increases in $r_{SE}$ (corresponding to an increasingly broad distribution of segmental mobilities), $G(t)$ exhibits an initial regime of more rapid reduction in $G(t)$ with time. For intermediate values of $r_{SE}$, at short times the slope is $r_{SE}$ dependent; at long times the Rouse scaling of $G \sim t^{-1/2}$ is recovered, albeit with $G(t)$ offset downward relative to the bare Rouse prediction. For very large $r_{SE}$, the short-time scaling exponent saturates to -1. Under the common observation discussed in the introduction that lower temperatures are associated with greater dynamic heterogeneity[27], larger values of $r_{SE}$ are expected to correspond to lower temperatures for a given system.

The physical origin of this alteration in scaling exponent is as follows. We can recast equation (6) in terms of the ratio of the effective mean segment friction coefficient $\langle\zeta_0\rangle_{eff,n}$ that is probed by a strand of $n$ repeat units to the true first moment mean segmental friction coefficient $\langle\zeta_0\rangle$:

$$\frac{\langle\tau_n\rangle}{\langle\tau_0\rangle} = n^2\left[1 - \left(1 - \frac{1}{n}\right)\left(1 - \frac{1}{r_{SE}}\right)\right] = n^2\left[\frac{\langle\zeta_0\rangle_{eff,n}}{\langle\zeta_0\rangle}\right] \quad (10)$$

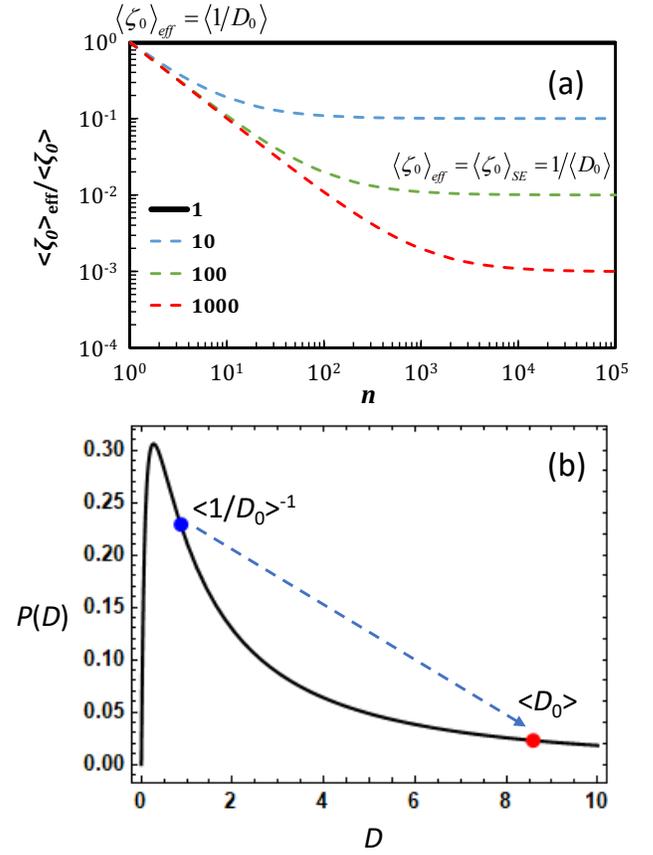

Figure 2. (a) Ratio of the effective friction coefficient probed by relaxation of a strand of length $n$ to the first moment of the chain friction coefficient distribution, plotted vs strand length. Inset equations note the moment of an underlying distribution of segmental mobilities $D_0$ that is probed by short and long strands (left and right equations, respectively). (b) Distribution of segmental mobilities $D_0$ (where locally $D_0 = 1/\zeta_0$) when $r_{SE} = 10$. Blue and red points in the figure note one over the inverse moment of this distribution and its first moment, with chain dynamics interpolating between a reflection of these two averages as chain length is increased.

As shown by Figure 2a, this ratio drops with increasing chain length when $r_{SE} > 1$, before ultimately saturating to a fixed value. Because short times in Figure 2a sample short-strand relaxation, this initial drop in the effective friction coefficient causes the 'compression' (increase in the magnitude of the power law exponent) at short times in Figure 1a. Figure 2b illustrates the physical origin of this drop, using as an example a case where $r_{SE} = 10$. With increasing strand length (the relaxation of which is sampled at increasing time), a strand interpolates between sampling the inverse moment of an underlying segmental mobility distribution function and sampling its first moment. This sampling obeys central limit theorem sampling, which leads to the $1/n$ scaling in the second term of equations (6) and (10). For $n \cong r_{SE}$, a single strand fully samples the segmental mobility distribution, and longer strands thus sample an *n*-invariant effective friction coefficient. This crossover corresponds to the return to Rouse scaling observed for long times Figure 1a and b, where relaxation samples length $n > r_{SE}$. It follows that this crossover occurs at $t/\tau_0 = r_{SE}$.

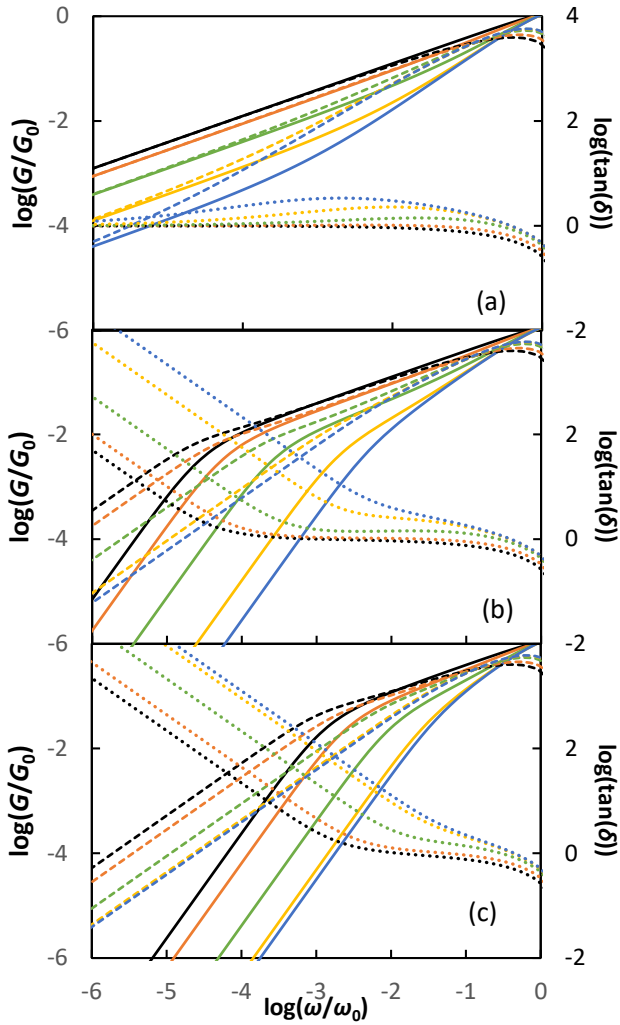

Figure 3. HRM predictions for log-log plots of storage modulus (solid lines, left axis), loss modulus (dashed lines, left axis), and tan(δ) (dotted lines, right axis) vs frequency, where frequency is normalized by the segmental frequency and moduli by their value at this frequency. Plots are for hypothetical unentangled chains of (a) $n \to \infty$, (b) $n = 200$, and (c) $n = 30$. Curves in each data set correspond to $r_{SE} = 1$ (Rouse model prediction, black line), $r_{SE} = 2$ (orange line), $r_{SE} = 10$ (green line), $r_{SE} = 100$ (yellow line), and $r_{SE} = 1000$ (blue line). For the storage and loss moduli, these are ordered from top to bottom; for tan(δ) they are ordered bottom to top.

As shown by Figure 1 c-f, systems with realistic subentangled molecular weights are unlikely to exhibit a fully developed exponent of -1 with a crossover to Rouse; the maximum number of decades of relaxation time that can be spanned by the Rouse regime in a real entangling polymer is insufficient. Instead, in realistic systems the HRM correction to Rouse corresponds to an apparent enhancement in the magnitude of the (negative) apparent power law slope in the 'Rouse' regime, with this enhancement again reflecting progressive sampling with increasing strand length of the distribution of segmental relaxation times. Thus, for the subentangled molecular weights for which the HRM fully applies, the relaxation modulus is predicted to *approximately* evolve as

$$\frac{G(t)}{G_0} \approx \tilde{t}^{\lambda(r_{SE})} \exp\left(-\frac{\tilde{t}}{\tau_n}\right) \quad \tilde{t} > 1 \qquad (11)$$

Where $\lambda(r_{SE})$ is an *effective* scaling exponent in the HRM regime at low molecular weight that is dependent on the degree of dynamic heterogeneity parameter $r_{SE}$. As shown above, $\lambda(r_{SE})$ reverts to ½ when $r_{SE} \to 1$ (single-valued segmental mobility) and goes to 1 as $r_{SE} \to \infty$ (extremely broad distribution of segmental mobilities). We emphasize that the true slope predicted by the HRM in the Rouse regime is time dependent, and that λ and equation (11) are simply convenient empirical approximations for short chains.

Experimental work aiming to quantify behavior in the Rouse regime near $T_g$ has generally employed measurements of either creep compliance or of the complex modulus. Since these functions are fully determined by $G(t)$, in the linear regime, they include no additional physical content relative to $G(t)$ predictions. Here we compute the complex modulus via a Fourier transform[53] of equation (8) in order to facilitate comparison to experimental complex modulus data. This transform cannot be analytically evaluated for the HRM $G(t)$ functional form (equation (8)), and we therefore perform the transform numerically.

In Figure 3a, we illustrate the predicted behavior of the complex modulus for the conditions shown in figure 1a and b: hypothetical unentangled chains of infinite molecular weight. As can be seen here, the behavior of the storage modulus mirrors that of $G(t)$ in this limit: the slope of the Rouse-like regime is enhanced at high frequencies with increasing heterogeneity, with Rouse scaling recovered at low frequency albeit with a downward shift of $G'(\omega)$. The corresponding tan(δ) develops a peak in the Rouse regime with increasing heterogeneity. For finite chains (Figure 3b and Figure 3c), these core features are retained, but the return to Rouse scaling at low frequency is lost due to truncation by the terminal mode – a result consistent with the findings of $G(t)$.

These predicted features are qualitatively consistent with previously reported experimental complex moduli near $T_g$, including results of Inoue et al, who report complex moduli for a range of temperatures for low molecular weight polystyrene[18], and of Cavaille et al, who report similar findings at higher molecular weight[54]. Inoue et al.'s data exhibit an enhancement in slope of the storage modulus in the Rouse regime on cooling, a shift of the terminal regime to high frequencies with decreasing temperature, and a growth in tan(δ) in this regime.

Cavaille's data agree on all counts. These features are all predicted by the HRM, as shown in Figure 3.

## Experimental Methods

In order to allow for more quantitative comparison to consistent experiments across multiple polymers, we perform small-amplitude oscillatory shear experiments and obtain $G^*(\omega)$ for three polymers. We specifically study poly(2-vinyl pyridine) (P2VP), poly(methyl methacrylate (PMMA), and polystyrene (PS). All three polymers we obtained from Scientific Polymer Products and used as obtained. We focus on nearly monodisperse polymers near or just below their entanglement molecular weight (see properties in Table 1), so as to enable study of the broadest possible Rouse regime without introducing entanglement effects.

Table 1. Properties of polymers studied. Molecular weight $M_n$ and polydispersity index are reported as indicated by the supplier, while $T_g$ was measured by DSC.

|       | $M_n$ | PDI  | $T_g$ / °C |
|-------|-------|------|------------|
| P2VP  | 8700  | 1.05 | 85         |
| PMMA  | 9120  | 1.03 | 120        |
| PS    | 12000 | 1.01 | 92         |

Polymer glass transition temperatures were determined via dynamic scanning calorimetry (DSC), performed on TA DSC2500. Glass transition temperatures $T_g$ were determined from heat flow curves at a cooling rate of 10K/min via the Moynihan equal-area method.

Complex moduli measurements were performed on an Anton Paar MCR 102 rheometer, employing a 4mm parallel-plate measurement geometry. Powder polymer samples were loaded into this instrument, heated above their respective glass transition temperatures (to a temperature of 150 °C for P2vP and PS and to 170 °C for PMMA), and held at this temperature for approximately 20 minutes. Each sample is then subject to a downward temperature sweep, taking rheometric measurements at intervals of 5K. For PS, data were collected over the range 130 °C-70 °C; for PMMA were collected over the range 160 °C-80°C,; for P2VP, data were collected over the range 130°C-95°C. Prior to the measurement at each temperature, we let the sample equilibrate under the target temperature to achieve a tolerance within 0.2 K for 180s. P2VP complex moduli were collected at each temperature over a frequency range of 0.01 rad/s-100 rad/s, and PS and PMMA complex moduli were collected at each temperature over a frequency range of 0.1 rad/s to 100 rad/s. Viscosity measurements were performed on the same instrument and geometry, at a low shear rate in the vicinity of 0.0015/s.

Dielectric relaxation was measured for the P2VP sample, employing the Solartron ModuLab XM MTS for dielectric measurements and an Anton-Parr MCR 702 with as a temperature-controlled sample stage. Two 25mm diameter parallel sample plates, with gap thickness of 100 μ set by the instrument, were used to conduct the measurement. Samples were loaded 180°C and held for 20 min, after which dielectric spectra were collected over the temperature range of 180°C to 85°C. Dielectric response was collected at each temperature over a frequency range of 100khz to 1hz.

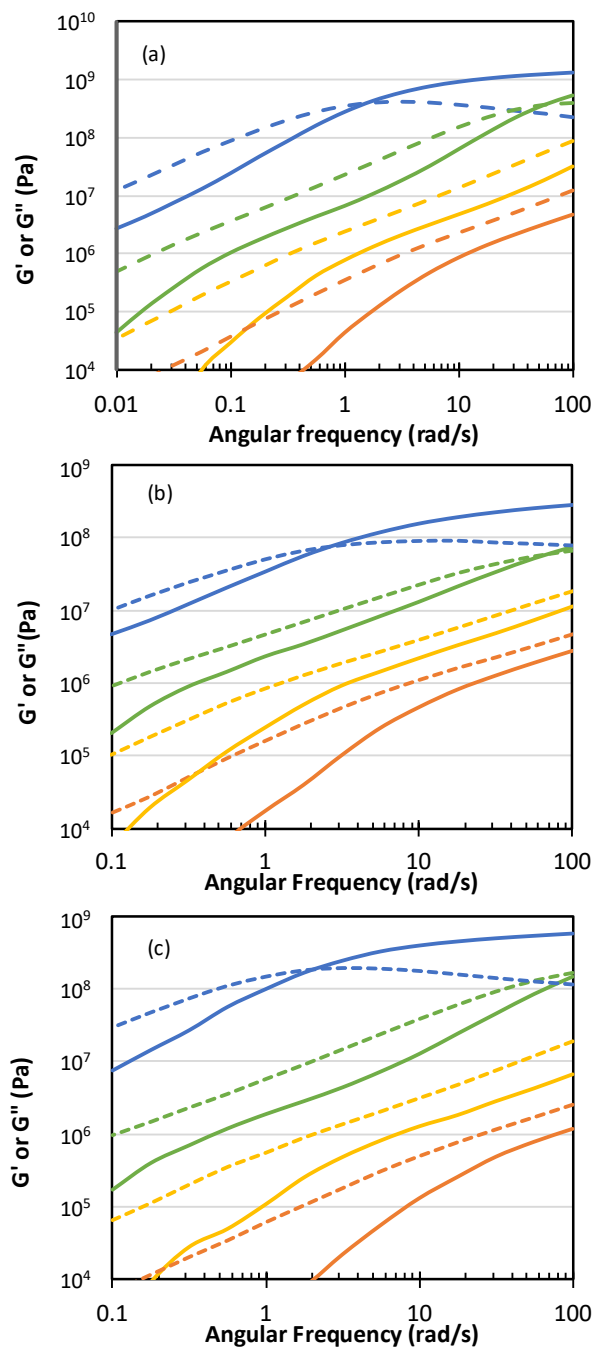

Figure 4. Representative complex moduli curves for (a) P2VP, (b) PMMA, and (c) PS. Solid lines are storage moduli and dashed lines are loss moduli.

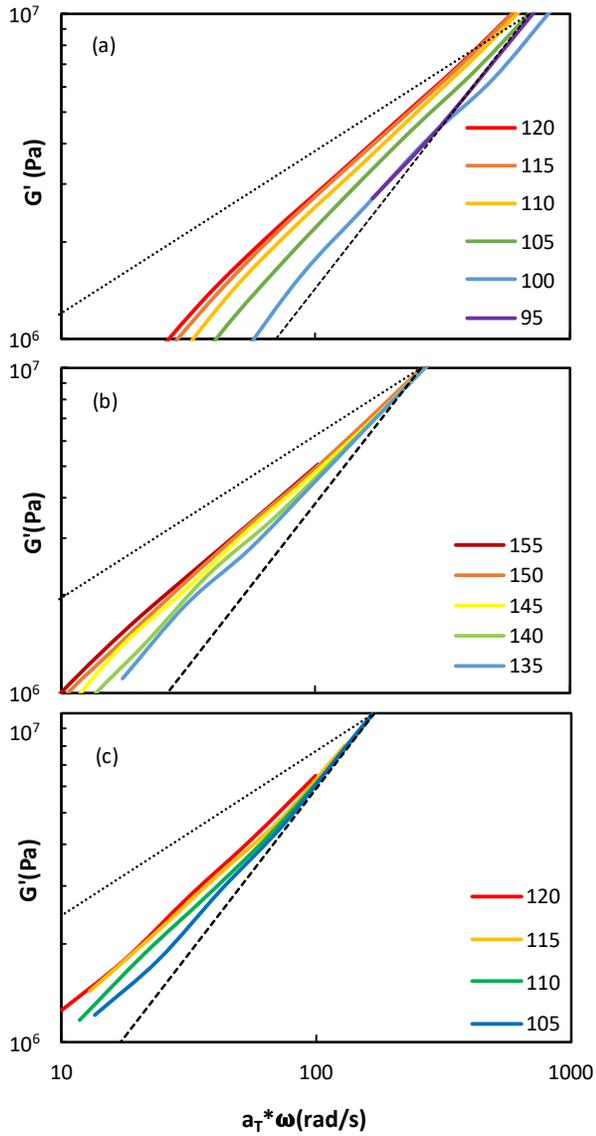

Figure 5. Shifted storage moduli, zoomed into the Rouse regime, for (a) P2VP, (b) PMMA, and (c) PS.

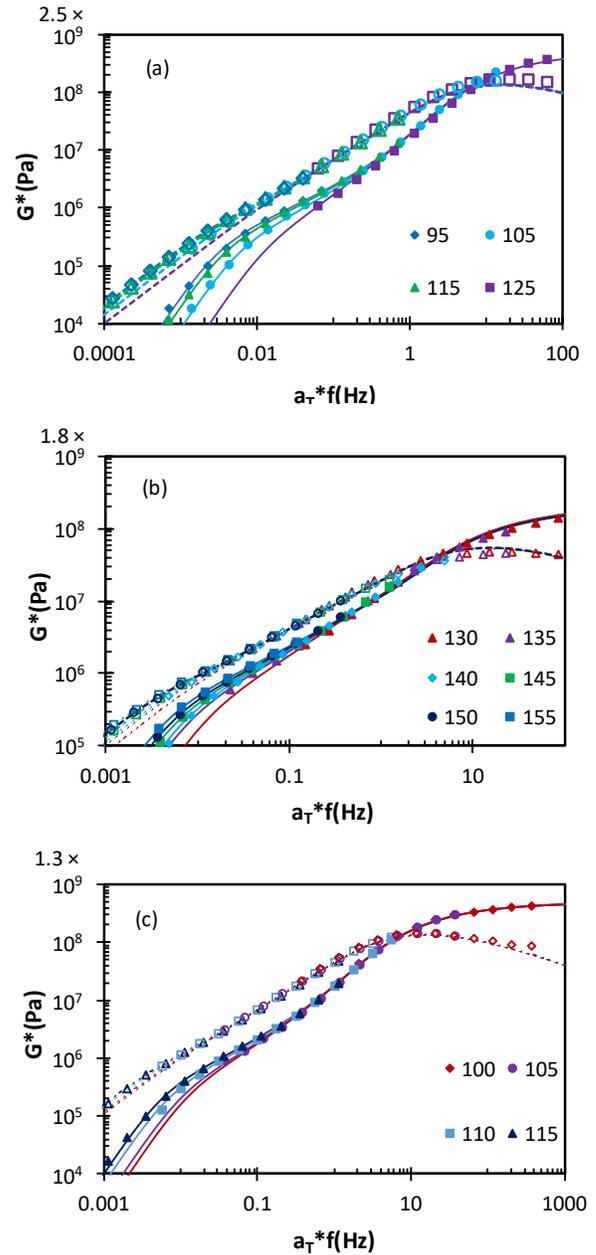

Figure 6. Shifted storage (filled points) and loss (open points) moduli for (a) P2VP, (b) PMMA, and (c) PS, for the temperatures shown in the captions. The solid curves represent HRM descriptions of the data at these temperatures.

## Results and Comparison to Theory

### Complex moduli

Representative complex moduli obtained for P2VP, PMMA, and PS are plotted in Figure 4. We focus on a range of temperatures for each system that encompasses the glassy relaxation, the Rouse regime, and the transition to terminal relaxation. Within these figures, the Rouse-regime generally corresponds to the modulus range of approximately $10^6$ Pa to $10^7$ Pa. In Figure 5, we zoom into this regime after applying a shift factor for each temperature to approximately superpose the data at the onset of the Rouse regime. This plot reveals two key features of the Rouse regime in all three systems. First, the slopes are consistently intermediate between the expected Rouse slope of ½ and a value of 1. Second, the slope generally grows on cooling. This behavior is qualitatively consistent with the predictions of the HRM described above, which anticipates a growth of slope in the Rouse regime on cooling as the extent of dynamical heterogeneity increases.

To more quantitatively compare the HRM predictions to these data, we introduce an empirical description of the segmental relaxation process at times $t < \tau_o$ as a stretched exponential process, such that the overall relaxation modulus as described as follows:

$$\frac{G(t)}{G_0} = \frac{G_{glass}}{G_0} \exp\left[-\left(\frac{t}{\tau_\alpha}\right)^\beta\right] +$$

$$\left(1 - \exp\left[-\frac{t}{\tau_0}\right]\right)\left[1 - r_{SE}\sqrt{\frac{1 + 2r_{SE}\left(2\frac{t}{\tau_0} - 1\right)}{+ r_{SE}^2}}\right]^{-1} 2\exp\left(-\frac{t}{N\tau_0}\frac{r_{SE}}{r_{SE} - 1 + N}\right) \quad (12)$$

We then obtain the complex modulus via numerical Fourier transform as before. The basic physical notion here is that the segmental relaxation process first decays, followed by the spectrum of heterogenous Rouse mode decays modeled by the HRM. We seek only a rough approximate description of the alpha relaxation process given that it is not the subject of the HRM model: we choose $\tau_\alpha$ such that the segmental process has essentially fully decayed by $\tau_o$ (i.e. decayed to $e^{-4.6}$ at $t = \tau_o$), and we employ a fixed stretching exponent $\beta$ equal to 0.5 for all systems, chosen as a reasonable value for diverse polymers). We find that these values provide a reasonable empirical description of the glass-to-Rouse crossover regime dominated by segmental α relaxation for all systems. We thus emphasize that the comparison to experiment in the $\omega > \omega_o$ regime is purely empirical, and only the regime wherein $\omega < \omega_o$ is relevant to assessing the physical predictions of the HRM.

As shown by Figure 6, the HRM provides an excellent quantitative description of both storage and loss in this regime, including the modified Rouse regime, for all three systems studied. For each polymer, the data are well-described with $G_{glass}$, $G_o$, and n independent of temperature, as expected by the HRM model, with values shown in Table 2. $r_{se}$ is varied with temperature to track the temperature-evolution of the slope in the Rouse regime.

Table 2. Characteristic pre-Rouse modulus $G_0$, ratio of glassy to pre-Rouse moduli, and number of dynamical repeat units employed in the HRM model to describe each experimental system studied.

|       | $G_o$             | $G_{glass}/G_o$ | n  |
|-------|-------------------|-----------------|----|
| P2VP  | $2.5 \times 10^7$ | 50              | 46 |
| PMMA  | $1.8 \times 10^7$ | 20              | 35 |
| PS    | $1.3 \times 10^7$ | 50              | 35 |

### Temperature dependence of decoupling

The temperature evolution of $r_{se}$ found to describe the data is reported in Figure 7. As can be seen here, $r_{se}$ is found to grow on cooling, which is consistent with the expectation that dynamic heterogeneity should grow on cooling given the equation (5) relationship between $r_{SE}$ and the breadth of the relaxation rate distribution. At equal $T_g/T$, PS and PMMA are observed to exhibit similar values of $r_{se}$, while P2VP exhibits somwehat greater $r_{SE}$.

Prior literature has suggested that trends in chain-segment decoupling and in dynamic heterogeneity may track with the kinetic fragility of glass formation, although empirically such trends appear to be fairly noisy at best[10]. We probe this proposed relationship in our systems by obtaining a calorimetric kinetic fragility index. To do so, we vary the cooling rate during calorimetry and measure the dependence of $T_g$ on cooling rate $r$, shown in Figure 10. The kinetic fragility index is generally defined as

$$m = \left.\frac{d\log\tau}{dT_g/T}\right|_{T=T_g}, \quad (13)$$

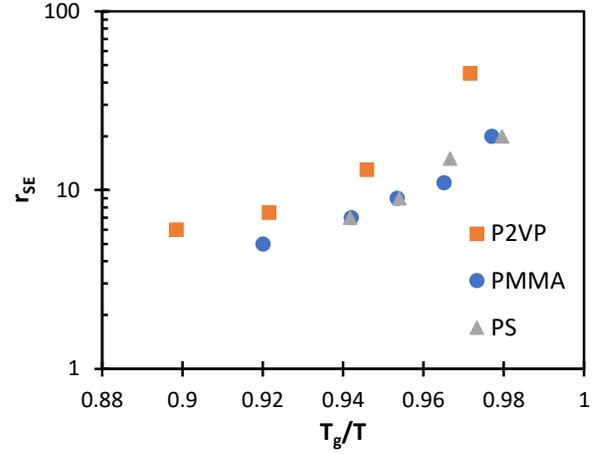

Figure 7. Stokes-Einstein breakdown parameter $r_{se}$ employed in the HRM models for the experimental systems shown in the legend, plotted vs $T_g$-normalized inverse temperature.

We obtain $m$ from the data above by fitting an Arrhenius rate law to the rate vs $T_g$ data. Combination of an Arrhenius rate law with equation (13), employing the relationship that at $T_g$ $r \propto \tau^{-1}$, yields the equation

$$m_{\text{calorimetric}} = \frac{E_A}{k_B T_g^{10K/min} \ln 10} \quad (14)$$

where $E_A$ is the fit Arrhenius activation energy and $k_B$ is Boltzmann's constant. This yields $m_{\text{calorimetric}} = $ 48, 76, and 52 for P2VP, PMMA, and PS, respectively. Evidently the modest difference in dynamic heterogeneity inferred from the HRM description of these systems is not readily predicted from $m_{\text{calorimetric}}$.

We note that the fragilities inferred here from calorimetry appear to be somewhat less than those previously reported for some of the systems from dielectric spectroscopy in a similar molecular weight range[11]. To confirm that our systems are nevertheless consistent with those prior studies, we perform broadband dielectric spectroscopy on our P2VP sample. We obtain dielectric spectra, as shown in Figure 10, and we extract an approximate relaxation time as the peak time of the alpha process (the higher-frequency of the two processes observed; the lower-frequency process may relate to the recently reported "Slow Arrhenius Process",[55] and is not the subject of the present study). As shown in Figure 8, we then fit the temperature-dependent relaxation times to the MYEGA model,

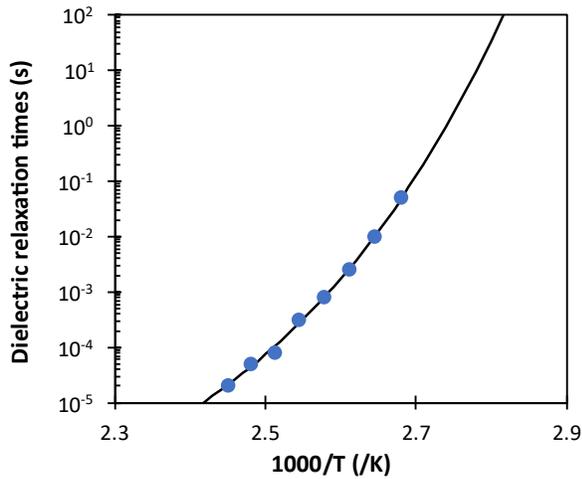

Figure 8. Dielectric relaxation time (points) and fit to MYEGA model (curve) vs inverse temperature.

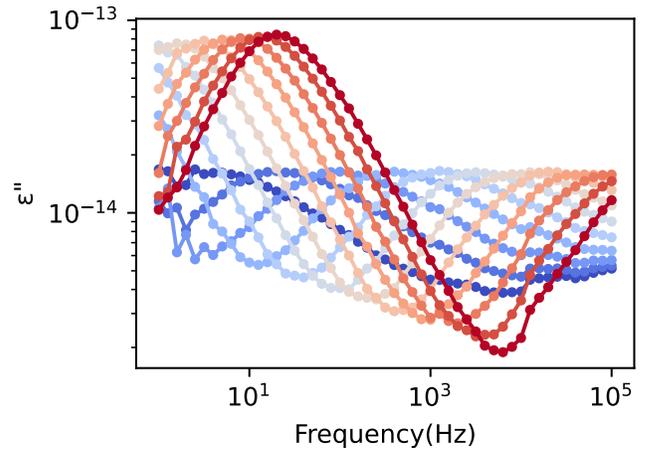

Figure 9. Dielectric loss spectra for P2VP, with curves at 5K intervals ranging from 85 °C (darkest blue) to 140°C (darkest red). The segmental relaxation process is the weaker process present at higher frequencies.

$$\ln \tau = \ln \tau_\infty = \frac{A}{T}\exp(-BT) \quad (15)$$

which we have previously shown to perform well in yielding reasonable determines of fragility and $T_g$ from data at somewhat shorter timescales or higher frequencies[56]. This analysis yields a dielectric fragility $m_{dielectric} = 83$ for P2VP – a value considerably higher than the calorimetric fragility reported above. This value, however, is consistent with prior determinations of the fragility of the dielectric alpha process in this molecular weight range[11].

This difference emphasizes the fact that fragility values can be quite different for distinct process, even when they are both putatively associated with segmental relaxation. Fragility values can also differ appreciably based on details such as the fitting function employed to extract a fragility from relaxation time data[56]. These factors may play an important role in the significant noise often found in fragility relationships, such as those between chain-segment decoupling and fragility[10]. This may also be consistent with other contexts, such as the strength of nanoconfinement effects on the glass transition, wherein fragility has been found to be at best a leading-order predictor of other trends in glass formation behavior[57]. Thus, while fragility apparently does exhibit some correlation with chain normal mode decoupling (and thus implicitly with dynamic heterogeneity via the HRM) as shown in prior work, it is evidently not fully predictive in any practical sense.

### Decoupling of viscosity and chain relaxation times

The HRM evidently provides an excellent description of experimental linear rheological data in these polymeric systems. We thus turn back to the remaining major objection to such heterogeneity-based explanations of TTS-breakdown: the observation that, in some systems, chain diffusion and shear viscosity can be decoupled in the presence of TTS-breakdown[5,16]. In contrast, in a lower molecular weight polymer, diffusion and viscosity were observed to be reasonably well coupled[58]. The essential challenge here is the intuition that, if averaging over dynamical heterogeneity is the origin of chain mode decoupling, then all relaxation times associated with whole-chain motion should be coupled given that they all average over the same size scale. However, this intuition has never been tested via a quantitative theory of the impacts of heterogeneity on chain motion, and it is thus not clear whether this behavior is generally at odds with the heterogeneity hypothesis for TTS breakdown. Moreover, it is not clear why this effect might vary between distinct systems or molecular weights and whether this variation could also be consistent with a heterogeneity origin of TTS-breakdown.

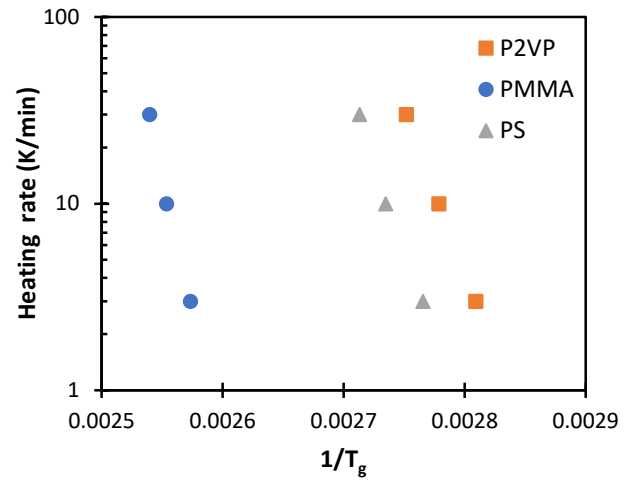

Figure 10. Ellipsometric heating rate vs inverse ellipsometric $T_g$ for the three experimental systems studied, as shown in the legend.

We thus perform a final HRM theoretical calculation to determine whether this quantitative implementation of the heterogeneity scenario for TTS-breakdown predicts decoupling between distinct chain-level processes. To do so, we combine the HRM with the common empirical observation of a fractional power law dependence of decoupling on relaxation time scale (equation (1)). Specifically, we define a 'fundamental' decoupling relation at the level of Einstein-breakdown,

$$r_{SE} \equiv \frac{\langle \zeta_0 \rangle}{\langle \zeta_0 \rangle_{SE}} \cong \frac{\langle \tau_0 \rangle}{\langle \tau_0 \rangle_{SE}} = \left(\frac{\langle \tau_0 \rangle}{\tau^*}\right)^\varepsilon \quad (16)$$

where $\tau^*$ is a timescale constant and $\varepsilon$ is called the "decoupling exponent", and where this is consistent with the observation of a fractional power law decoupling relation between diffusion and relaxation in small molecule systems[59]. Combination of equations (6) and (16) leads to the following equation for the chain relaxation time as a function of the mean segmental relaxation time.

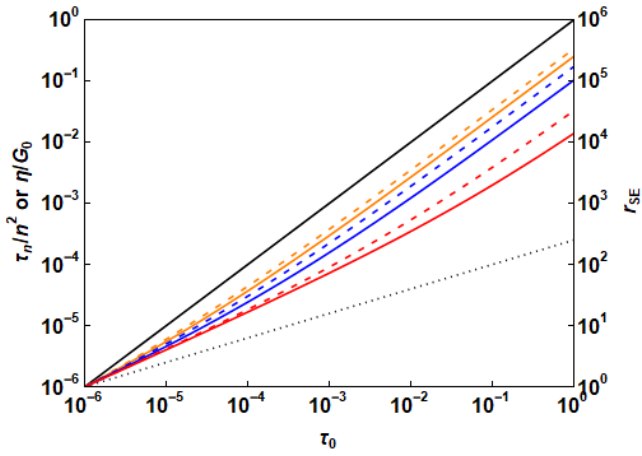
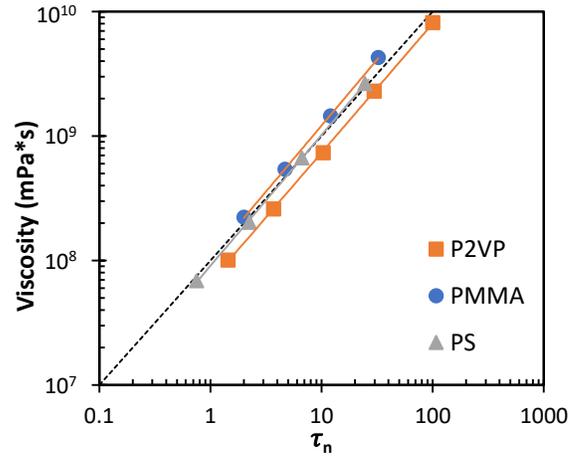

Figure 11. Normalized (as shown on the left y-axis) chain relaxation time (solid lines) and viscosities (dashed lines) plotted vs segmental relaxation time, with these values computed as described in the text based on an underlying decoupling relation (equation (16)) with $\tau^* = 10^{-6}$ s and $\varepsilon^* = 0.4$ (the approximate value for polystyrene). The solid black line is the Rouse prediction for both quantities on the left axis. Other data sets from top to bottom are for chain lengths of $n = 4$ (orange), $n = 10$ (blue), and $n = 100$ (red). All axes are in units of seconds. The dotted line, which is the only data set associated with the right y-axis, plots $r_{SE}$ vs $\tau_o$ for the combination of $\tau^*$ and $\varepsilon$ shown above. This is the underlying trend *assumed* in this calculation to obey the fractional power law form as an empirical matter. In other words, the dotted line is an input to the model predictions in this figure, not a prediction.

Figure 12. Viscosity vs approximate terminal relaxation time for each of the systems studied, as shown in the legend. Points are experimental data. Solids lines are power law fits to the data. The dashed line is a power law with scaling exponent one, for visual comparison.

$$\langle \tau_n \rangle = \langle \tau_0 \rangle n^2 \left[1 - \left(1 - \frac{1}{n}\right)\left(1 - \left(\frac{\tau^*}{\langle \tau_0 \rangle}\right)^\varepsilon\right)\right] \quad (17)$$

By inserting equation (16) into the relaxation modulus and integrating, we can also arrive at a prediction for the viscosity as a function of segmental relaxation time. To do so, we must select values of $\varepsilon$ and $\tau^*$. Data in the literature suggest that polymers exhibit a range of $\varepsilon$ values from about 0 to 0.6.[10] The constant $\tau^*$ sets the segmental relaxation time timescale beyond which (temperature below which) heterogeneity begins to grow and segmental and chain modes thus become decoupled. Based on literature data, $\tau^*$ appears to range from approximately 1 μs to 1 ms over a set of several common polymers exhibiting varying degrees of decoupling[10,12].

In Figure 11 we plot HRM predictions for chain relaxation times and viscosities as a function of segmental relaxation time in the case where $\varepsilon^* = 0.4$ and $\tau^* = 1$ μs. First, this figure makes clear that both chain relaxation and viscosity are decoupled from segmental relaxation. However, as can be seen here, *the HRM also predicts that chain relaxation times and viscosity are weakly decoupled, despite both reflecting chain-scale motion*. This prediction essentially amounts to an expectation that, when segment and chain motion is decoupled due to dynamic heterogeneity, chain relaxation and viscosity should be weakly decoupled as well. Moreover, the theory predicts the direction of this decoupling, predicting that viscosity should have a modestly stronger temperature dependence than whole-chain relaxation time.

To test whether this prediction holds for our data, we perform viscosity measurements as a function of temperature on our experimental systems. Since our systems are unentangled, there is no pronounced peak in tan(δ) at the terminal (chain) relaxation frequency. We therefore extract an approximate chain relaxation frequency $\omega_n$ as the frequency at which $G'(\omega) = 10^5$, which as can be seen in Figure 4 corresponds visually to the approximate onset of terminal behavior. We then define a chain relaxation time $\tau_n = 1/\omega_n$.

To determine whether η genuinely exhibits a systematically larger temperature dependence than $\tau_n$, in Figure 12 we plot η vs $\tau_n$. As can be seen in this figure, the slopes of the η vs $\tau_n$ relationship appear to be marginally greater than one (note, for example, the crossover between the line of slope 1 and the fit curve for PS), consistent with the mild decoupling predicted between these quantities by the HRM. To better quantify this decoupling, we perform a fit of these data to the fractional power law decoupling relation form. These fits all yield a decoupling exponent $\varepsilon_{\eta\text{-}\tau n}$ for the η vs $\tau_n$ relationship modestly greater than one, with $\varepsilon_{\eta\text{-}\tau n}$ equal to 1.040 ± 0.0230, 1.061 ± 0.310, and 1.052 ± 0.0650 for P2VP, PMMA, and PS respectively (and where the confidence intervals are at the 95% level for the fit).

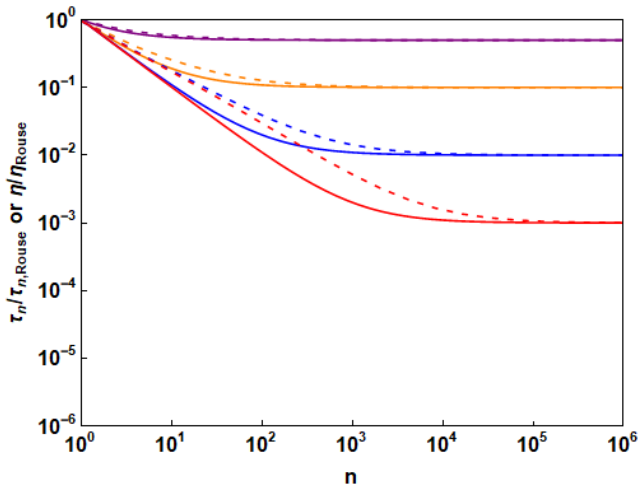

Figure 13. Ratios of chain relaxation time (solid) and viscosity (dashed) to their values when $r_{SE} = 1$ (i.e. the Rouse model), plotted vs chain length, for values of $r_{SE}$, from top to bottom, of 2 (purple), 10 (orange), 100 (blue), and 1000 (red). Calculations are for hypothetical non-entangling chains, which become aphysical for $n$ of order appreciably larger than 100. Note that data sets here have no direct relation to those in Figure 11 since data sets here correspond to fixed values of $r_{se}$ while those in Figure 11 correspond to fixed values of $n$.

The finding of $\varepsilon_{\eta\text{-}\tau n} > 1$ in all three systems is consistent with and confirms the HRM prediction of a stronger temperature dependence of viscosity than chain relaxation times.

Notably, this decoupling between chain relaxation and viscosity is predicted to become more pronounced with increasing chain length. This may explain the puzzling apparent point of disagreement in the literature discussed in the introduction. As noted by Ngai and coworkers, this type of decoupling between chain motion and viscosity has been observed in some *entangled* polymers[5,16]. On the other hand, Urakawa, Swallen, Ediger, and von Meerwal[58] have reported that the self-diffusion coefficient and viscosity of *low-molecular-weight* polystyrene (of order a few thousand g/mol) are nearly coupled, at least to within one decade. The HRM predictions in Figure 11 point to a potential resolution of this apparent conundrum: longer chains exhibit a larger degree of decoupling between chain relaxation and viscosity – at least over the range n=4 to n=100, which for many polymers covers nearly the entire range from oligomer to entanglement strand.

The physical origin of this chain-level decoupling can be readily understood within the HRM based on the approximate scaling of $G(t)$ given by equation (11). Specifically, the ratio of viscosity to terminal chain relaxation time is given by

$$\frac{\eta}{G_0 \tau_n} \equiv \frac{\int_0^\infty G(t)dt}{G_0 \tau_n} = \frac{\tau_0}{\tau_n}\left[1 + \int_1^\infty \frac{G(\tilde{t})}{G_0} d\tilde{t}\right]$$
$$\cong \frac{\tau_0}{\tau_n}\left[1 + \int_1^\infty \tilde{t}^{-\lambda} \exp\left(-\frac{\tilde{t}}{\tau_n}\right) d\tilde{t}\right] \quad (18)$$

For small $n$, this quantity is equal to 1. With increasing $n$, it progressively drops, effectively introducing an $n$-dependent conversion factor between viscosity and chain terminal relaxation time; this is well known and is the origin of the fact that chain relaxation time scales as $n^2$ while the viscosity scales as $n$ within the Rouse model. Within the Rouse model, the ratio of these scaling exponents is temperature-independent. However, the HRM predicts that this ratio depends on $r_{SE}$, and thus on temperature as indicated by the temperature evolution of $r_{se}$ in the experimental data reported above. This is a consequence of the distorting of the Rouse regime via the trend in the effective exponent $\lambda$ as $r_{SE}$ grows on cooling. Again, this can be tracked directly back to the effect of intrachain 'pre-averaging' over a growing distribution of relaxation times.

This effect, its dependence on chain length, and its physical origins can also be understood from Figure 13, which plots chain relaxation times and viscosities, each normalized by their Rouse ($r_{SE} \to 1$) predictions, as a function of chain length. As can be seen here, chain relaxation times and viscosities exhibit *qualitatively* comparable behavior: with increasing $n$, their values drop below the Rouse prediction due to intra-chain averaging over dynamic heterogeneity; at high $n$, the effect saturates once the entire distribution of segmental mobilities is averaged out within a single typical chain.

As per equation (18), in the low-$n$ limit viscosity and chain relaxation time track apart with increasing $n$ as a consequence of the growth of the domain of time controlled by the $r_{SE}$-dependent effective exponent $\lambda$. For hypothetical unentangled chains at *very* large $n$, well over the typical chain length of entanglement, viscosity and chain relaxation time are predicted to become coupled once again. This corresponds to the hypothetical high-$M$ regime shown in Figure 1 and Figure 3 wherein Rouse scaling of $G(t)$ is recovered (with a downward offset) at long times; in this regime the $r_{SE}$-dependent value of $\lambda$ at logarithmically short times is ultimately forgotten in the integral in equation (18) and terminal relaxation time and viscosity therefore once again track together. However, *the deviations from coupling between viscosity and chain terminal relaxation time become most pronounced in the n=10 to n=1000 range that is typical of subentangled polymers and of entanglement strands in entangled polymers.*

Ultimately, the HRM thus predicts that decoupling between chain-scale quantities is to be expected when TTS breaks down. This decoupling is generally expected to grow with chain length for strands up to typical entanglement lengths. This is a consequence of the fact that even quantities we commonly conceptualize as 'chain scale' can reflect modestly differing moments of the underlying relaxation time distribution. This prediction and associated experimental finding thus apparently resolves the final major objection reviewed in the introduction to a heterogeneity-based mechanism of TTS breakdown

## Discussion and Conclusions

Given the ongoing debate regarding the origin of TTS breakdown, we summarize here the physics, assumptions, and approximations that are encoded in the HRM model and on which our results rest.

1) The HRM model inherits all assumptions of the Rouse model, *other* than the assumption that friction coefficients are monodisperse.
2) The HRM model injects an (initially generalized) distribution of segmental friction coefficients into the Rouse model; this distribution is ultimately modeled as log-normal in form.
3) A leading-order mathematical treatment is employed in which dynamics are assumed to obey a Gaussian path process (as in the Rouse model) and in which the central limit theorem is employed to describe intra-chain averaging over the distribution described in (2).

The sets of assumptions above are well-supported empirically. In particular, the success of the Rouse model is well-supported by its long success in melts far from $T_g$. The inclusion of a distribution of segmental friction coefficients is well motivated by the literature, with numerous computational and experimental studies establishing that glass-forming liquids, including polymers near $T_g$, are generally dynamically heterogeneous, possessing a distribution of local mobilities[24–40]. The specific choice of a log-normal distribution is consistent with simulation and experimental studies indicating that this form describes the distribution of local mobilities or friction coefficients to leading order[25–27,52]; this choice is also consistent with a Gaussian distribution of local activation barriers and with an alternate version of the central limit theorem[50,51]. Finally, the success of this model in our prior work and here suggest that the mathematical leading order truncations described in point (3) are relatively small[46].

The three sets of assumptions 1-3 above *alone* lead to the modified scaling of friction coefficient and $\tau$ in $n$, and to the modified scaling of $G(t)$ and $J(t)$ in time and $G^*(\omega)$ and $J^*(\omega)$ in frequency. Given the parsimony of this model, it is fair to say that *the revised n- and t-scalings given above are what one should expect near $T_g$ based purely upon the empirical fact of the presence of a distribution of local mobilities*. In other words, these 3 elements lead to a prediction of modified Rouse scaling at $T \to T_g$. However, alone, they do not predict TTS-breakdown, since a temperature-invariant distribution of segmental friction coefficients within this model would lead to a temperature-invariant modified Rouse scaling that would still obey TTS.

In addition to points 1-3, an additional physical proposition is thus invoked to rationalize the *temperature-dependence* of deviations from Rouse scaling and resultant TTS breakdown:

4) The distribution of segmental friction coefficients broadens on cooling.

Again, considerable simulation and experimental evidence support this proposition[27,30,31,40,43–45]. However, this point is somewhat more contentious than the first three and has been a primary point of criticism of heterogeneity-based explanations for TTS-breakdown. Specifically, Ngai and coworkers have suggested that an absence of broadening in the alpha process of dielectric spectra of some fluids exhibiting fractional power law decoupling may contradict this scenario. However, as discussed in the introduction, our recent work offers a potential resolution to this apparent contradiction, suggesting that the breadth of dielectric response spectra may not be a reliable reporter of the breadth of the underlying spatial distribution of relaxation times[27,46].

Our rheological experimental data are well-described by the HRM predictions that emanate from the combination of points 1-4 above. We additionally resolve a second proposed objection to a heterogeneous origin of TTS breakdown - the observation of decoupling between distinct whole-chain processes, such as viscosity and chain relaxation time[16]. In particular, we show that this outcome is consistent with a heterogeneous origin of TTS-breakdown. This is a consequence of their still probing distinct moments of the overall relaxation process, despite all putatively reflecting 'whole chain' behavior.

Collectively, these findings indicate that known growth of dynamic heterogeneity upon cooling is sufficient to explain the emergence of thermorheological complexity upon homopolymers upon approach to $T_g$, with the Heterogeneous Rouse Model providing a predictive theory for rheological response in this regime. Furthermore, the prior results reviewed above, combined with our present findings, suggest that compression of the Rouse regime upon cooling may serve as a better probe of dynamic heterogeneity than does the stretching exponent for the segmental alpha process. Effectively, polymer chain normal modes serve as a natural probe of relaxation behavior averaged over distinct length scales. The HRM model provides a quantitative means of inferring the extent of dynamic heterogeneity from the Rouse compression that results. *This may provide a pathway toward a new approach to quantifying heterogeneity in polymeric glass-forming liquids.*

At important limitation of this model is that it predicts the *consequences* of emergent dynamic heterogeneity for chain relaxation and does not predict the emergence of dynamic heterogeneity itself. Such a prediction would require a segmental-scale theory of glass formation wholly distinct from the theory here. In this vein, the Elastically Cooperative Nonlinear Langevin Equation (ECNLE) theory of Mirigian and Schweizer[60–62] may provide more insight into the precise temperature dependence of this phenomenon. The ECNLE has recently been shown to predict heterogeneity-based fractional power decoupling in small-molecule liquids, in a manner that appears to be qualitatively consistent with the physics of the HRM[63]. Within the ENCLE theory, local density fluctuations (heterogeneity) alter both local caging and long-ranged elastic contributions to the activation barrier, leading to

Stokes-Einstein breakdown, the fractional power law form, and the approximate Gaussian distribution of activation barriers implied by the HRM usage of a log-normal distribution of local mobilities[63]. Given that the underlying ECNLE theory has been shown to provide good predictions for segmental relaxation in a range of polymers[62], future combination of the HRM with ECNLE predictions might have the potential to enable full prediction of polymer linear rheological response functions near $T_g$.

## AUTHOR INFORMATION

**Corresponding Author**

* dssimmons@usf.edu

**Funding Sources**

This material is based on work supported by the American Chemical Society Petroleum Research Fund New Directions Grant.

## Acknowledgments

DSS acknowledges helpful discussions with G. McKenna and K. Schweizer.